\pdfoutput=1

\documentclass{PoS}

\usepackage{amsmath}
\usepackage{amssymb}
\usepackage{cite}
\usepackage{epsfig}
\usepackage{latexsym}
\usepackage{booktabs}
\usepackage{graphicx}

\newcommand{\be}{\begin{equation}}
\newcommand{\ee}{\end{equation}}

\newcommand{\la}{\label}

\newcommand{\ba}{\begin{eqnarray}}
\newcommand{\ea}{\end{eqnarray}}
\renewcommand{\eqref}[1]{Eq.~(\ref{#1})}
\newcommand{\tabref}[1]{Table~\ref{#1}}
\newcommand{\figref}[1]{Fig.~\ref{#1}}

\newcommand{\ten}[1]{\times 10^{#1}}

\title{The pion quasiparticle in the low-temperature phase of QCD}

\ShortTitle{The pion quasiparticle in the low-temperature phase of QCD}

\author{Bastian B. Brandt\\
Institut f\"ur theoretische Physik, Universit\"at Regensburg, D-93040 Regensburg, Germany}

\author{Anthony Francis\\
Department of Physics \& Astronomy, York University, 4700 Keele St, Toronto, Canada}

\author{Harvey B.\ Meyer and \speaker{Daniel Robaina}\\
PRISMA Cluster of Excellence,
Institut f\"ur Kernphysik and Helmholtz~Institut~Mainz,
Johannes~Gutenberg-Universit\"at~Mainz,
D-55099 Mainz, Germany\\
        E-mail: \email{robaina@kph.uni-mainz.de}}

\abstract{We investigate the properties of the pion quasiparticle in the
low-temperature phase of two-flavor QCD on the lattice with support
from chiral effective theory.  We find that
the pion quasiparticle mass is significantly reduced compared to its
value in the vacuum, in contrast to the static screening mass, which
increases with temperature.  By a simple argument, the two masses are expected to determine the quasiparticle
dispersion relation near the chiral limit. Analyzing two-point functions of the axial
charge density at non-vanishing spatial momentum, we find that the
predicted dispersion relation and the residue of the pion pole are
simultaneously consistent with the lattice data at low momentum.  The
test, based on fits to the correlation functions, is confirmed by a
second analysis using the Backus-Gilbert method.}

\FullConference{The 8th  International Workshop on Chiral Dynamics \\
		 29 June 2015 - 03 July 2015\\
		 Pisa, Italy}

\begin{document}

\section{Introduction}
In order to obtain a deep understanding of the Quark-Gluon-Plasma medium and of its properties, it is of central importance to study in which way the individual excitations of the system are modified with increasing temperature. Viewed globally, the hadron spectrum does not appear to change very much compared to the zero temperature situation until $T\sim T_C$ where the transition to the deconfined phase takes place. This observation is based on the success of the Hadron Resonance Gas Model (HRG) in describing equilibrium properties of the medium, such as quark number susceptibilities for $T<T_C$. These predictions are in very good agreement with Lattice calculations (see e.g. \cite{bazavov}).

Nevertheless, detailed information about the individual degrees of freedom and of their dynamics at finite temperature is lacking. This information can be extracted from thermal Euclidean correlators which can be calculated at the percent level in Lattice QCD. Here, we extend our lattice study of \cite{ourref}, where we derived and tested a modified pion dispersion relation, \eqref{eq:dispersion_relation}, obtained from an effective theory, which is valid for $T \lesssim T_C$
\be
\omega_{\bf p} = u(T) \sqrt{m^2_\pi + {\bf p}^2}.
\la{eq:dispersion_relation}
\ee
The screening mass $m_\pi$ is defined by the inverse correlation length and $\omega_{\bf p}$ is the position of the pole on the $\omega$-axis in the retarded correlator $G_R(\omega, {\bf p})$ with pion quantum numbers. Under the assumption that one is sufficiently close to the chiral limit and for small values of the momentum, a single parameter $u(T)$ determines the full dispersion relation of soft pion quasiparticles at finite temperature (for details on the derivation of \eqref{eq:dispersion_relation} see \cite{ourref}). In the chiral limit $u(T)$ corresponds to the group velocity of a massless excitation $v_g = d\omega/d{\bf p} = u \le 1$ (in natural units). At $T=0$, SO(4) Euclidean symmetry implies $m_\pi=\omega_0$ and $u=1$.

In this proceedings article, we test the dispersion relation \eqref{eq:dispersion_relation} on a single thermal fine lattice ensemble below $T_C$ with high statistics and two dynamical degenerate light quarks \cite{ourref2}. In addition, we revisit an old method for inverting integral equations, the Backus-Gilbert method \cite{backus,backus2}, which, to our knowledge, is used here for the first time within QCD in order to obtain model independent estimators for spectral functions with input from Euclidean correlators. This is particularly important since real-time excitations of the system are encoded in spectral functions (e.g. via the hydrodynamic description and Kubo formulas \cite{teaney}).

\section{Lattice estimators of $u$ at ${\bf p}={\bf 0}$ \la{sec:lattice_estimators}}
By exploiting the parametric dominance of the pion in time-dependent axial-charge density correlators together with Ward Identities arising from the PCAC relation we derived two independent lattice estimators for the parameter $u(T)$ \cite{ourref} that are valid at any temperature below $T_C$ (see also \cite{stephanov,stephanov2}):
\begin{eqnarray}\la{eq:um}
u_m &=& \left[-\frac{4m^2_q}{m^2_\pi} \left.\frac{G_P(x_0,T,0)}{G_A(x_0,T,0)}\right|_{x_0 = \beta/2}\right]^{1/2},\\ 
u_f  &=& \frac{f^2_\pi m_\pi}{2 G_A(\beta/2, T, 0)\sinh(u_f m_\pi \beta/2)}.
\la{eq:uf}
\end{eqnarray}
The relevant correlation functions that enter the calculation are
\begin{eqnarray}
\la{eq:GA}
\delta^{ab}\,G_A (x_0, T, {\bf p}) &=& \int d^3x~ e^{i {\bf p\cdot x}}\left<A^{a}_0(x)A^{b}_0(0)\right>, \\
\delta^{ab}\,G_P (x_0, T, {\bf p}) &=& \int d^3x~ e^{i {\bf p\cdot x}} \left<P^a(x)P^b(0)\right>
\la{eq:GP}
\end{eqnarray}
where $a,b$ are flavor indices of the adjoint representation and $\beta=1/T$. The quark mass $m_q$ is defined via the PCAC relation \cite{PCAC}. The screening pion decay constant $f_\pi$ as well as the screening mass $m_\pi$ are defined through the long distance behavior of Euclidean correlators along spatial directions
\ba\la{eq:GAs}
\delta^{ab}\,G^\text{s}_{A} (x_3, T) &=& \int dx_0\; d^2x_{\perp} \left<A^{a}_3(x) A^{b}_3(0)\right> ~\stackrel{|x_3|\to\infty}{=}~ \delta^{ab} \frac{f^2_\pi m_\pi}{2}e^{-m_\pi|x_3|}, \\
\delta^{ab}\,G^\text{s}_{P} (x_3, T) &=& \int dx_0\; d^2x_{\perp} \left<P^a(x)P^b(0)\right> ~\stackrel{|x_3|\to\infty}{=}~  -\delta^{ab} \frac{f^2_\pi m^3_\pi}{8m^2_q}e^{-m_\pi|x_3|}.
\la{eq:GPs}
\ea
Notice that the parameter $u(T)$ is a RG invariant quantity (so are $u_m$ and $u_f$) and therefore renormalization constants cancel out. 

\section{Lattice setup}
Measurements where performed on a $24\times 64^3$ lattice with two dynamical light quarks with a mass of $m_q = \overline{m}^{\overline{\text{MS}}}(\mu=2\text{GeV}) = 12.8(1)\text{MeV}$. The temperature is $T=1/24a = 169 \text{MeV}$ and the spatial extent amounts to $L=64a=3.1\text{fm}$. In the chiral limit the critical temperature for $N_f=2$ is $T_C(0) \approx 170\text{MeV}$ \cite{TC} and since $T_C(m_q)$ grows with $m_q$ our thermal ensemble lies below the phase transition. In addition, an effective zero-temperature $128\times 64^3$ ensemble with the same bare parameters is available to us through the CLS effort (labelled as O7 in \cite{O7}). Therefore we are able to compare in a straightforward manner our finite temperature results with the vacuum situation. Results are shown in \tabref{tab:screening} and a comparison with previous work is shown in \figref{fig:pion_velocity}.

\begin{table}[h!]
\centering
\begin{tabular}{l@{~~~}r}
\hline \hline
$m_\pi/T$ & 1.79(2) \\ 
$f_\pi/T$ & 0.46(1) \\ \hline 
$u_f$ & 0.76(1) \\
$u_m$ & 0.74(1) \\ \hline
$\omega_{\bf 0}/T$ & 1.32(2)\\
$f_{\pi}^t/T$ & 0.62(1) \\ \hline \hline
\end{tabular}
\hspace{2cm}
\begin{tabular}{l@{~~~}r}
\hline \hline
$\omega_{\bf 0}/T$ & 1.579(12) \\
$\omega_{\bf 1}/T$ & 2.88(3)\\
$f^{0}_{\pi}/T$ & 0.599(8)\\ \hline
$u^2(T \simeq 0)$ & 1.01(6)\\ \hline \hline 
\end{tabular}
\caption{Left: Summary of the results for the $N_\tau=24$ thermal ensemble. The value of $\omega_{\bf 0}$ is calculated using $\omega_{\bf 0}=u_m m_\pi$. In the same way $f_{\pi}^t=f_\pi/u_m$. Right: Summary for the vacuum ensemble O7. The values of $\omega_{\bf 0}$ and $\omega_{\bf 1}$ are extracted from the long range behavior of \protect\eqref{eq:um} and \protect\eqref{eq:uf} with ${\bf p} = (0,0,2\pi n/L)$ and $n=0,1$ respectively. The value of $u^2(T \simeq 0)$ was extracted via a linear fit to the square of \protect\eqref{eq:dispersion_relation} with $m_\pi$ replaced by $\omega_{\bf 0}$. All renormalization factors are included and the errors are purely statistical. All dimensionfull results are normalized with $T=1/24a$.}
\label{tab:screening}
\end{table}

\begin{figure}[h!]
\begin{center}
\includegraphics[width=0.50\textwidth]{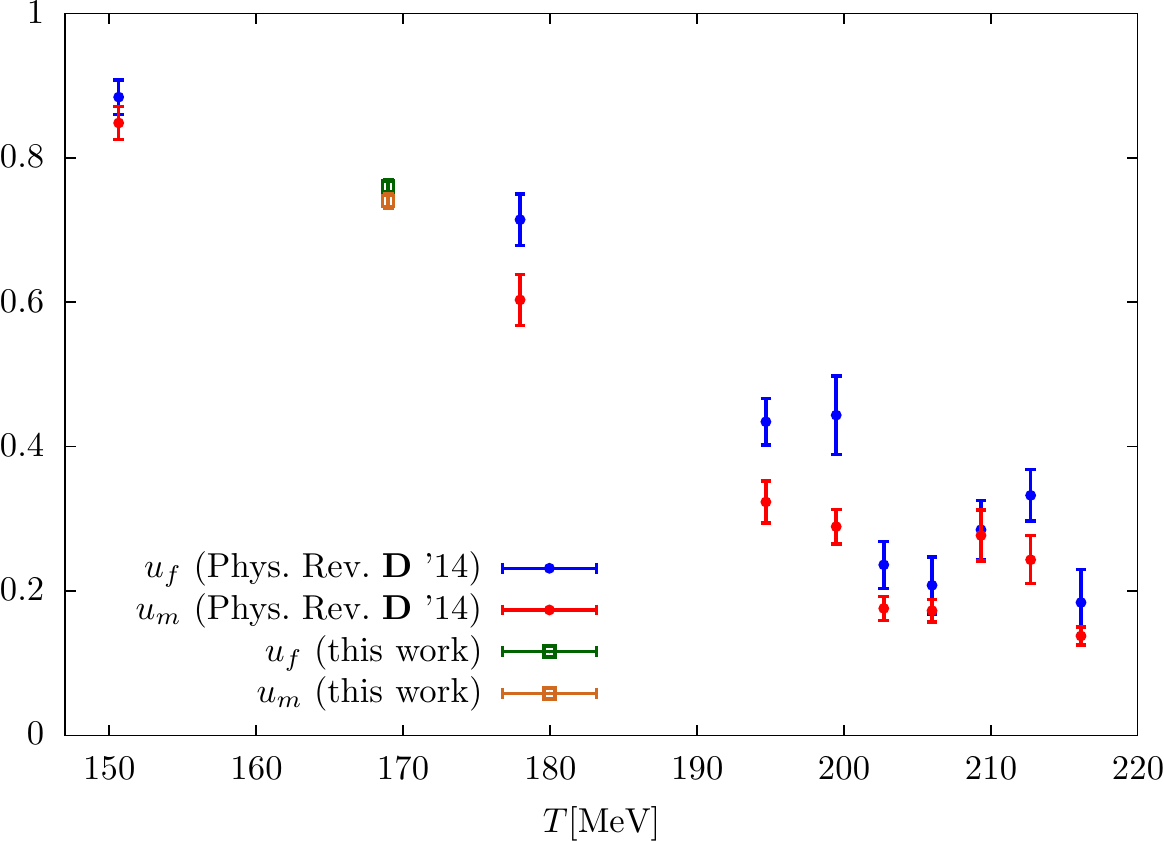}
\end{center}
\caption{Comparison of the results obtained in \cite{ourref} with this work \cite{ourref2}. Notice that in \cite{ourref} the temperature scan was performed at a constant renormalized quark mass of $\overline{m}^{\overline{\text{MS}}}(\mu=2\text{GeV}) \sim 15\text{MeV}$.}
\la{fig:pion_velocity}
\end{figure}

\section{Axial-charge density correlator at ${\bf p} \neq {\bf 0}$}
The relation between the Euclidean correlator $G_A(x_0, T, {\bf p})$ and its spectral function is
\be
G_A(x_0, T, {\bf p}) = \int^\infty_0 d\omega~ \rho^{A}(\omega, {\bf p}) \frac{\cosh(\omega(\beta/2-x_0))}{\sinh(\omega \beta/2)}.
\label{eq:spectral_def}
\ee
At finite temperature, the analysis of the correlator $G_A(x_0,T,{\bf p})$ is more involved than at zero temperature: only at sufficiently
small quark masses and momenta, and not too small $x_0$, is the
correlator parametrically dominated by the pion pole. Therefore, we
proceed by formulating a fit ansatz for the spectral function to take into account the non-pion
contributions with the property that $\rho^{A}(\omega \to \infty, T, {\bf p}) = \theta(\omega^2 - 4m^2 - {\bf p}^2) \frac{N_c}{24 \pi^2} ({\bf p}^2 + 6m^2)$ (see e.g. \cite{martinezresco}). By integration with the kernel $\frac{\cosh(\omega(\beta/2-x_0))}{\sinh(\omega \beta/2)}$ the spectral function and the correlator fit ans\"atze read
\ba
\rho^A(\omega, {\bf p}) &=& A_1({\bf p})\sinh(\omega\beta/2) \delta(\omega-\omega_{\bf p}) + A_2({\bf p}) \frac{N_c}{24\pi^2}\left(1-e^{-\omega \beta}\right) \theta(\omega-c),\\
G_A(x_0, T, {\bf p}) &=&  A_1({\bf p}) \cosh(\omega_{\bf p} (\beta/2-x_0)) + A_2({\bf p}) \frac{N_c}{24\pi^2} \left(\frac{e^{-cx_0}}{x_0}+\frac{e^{-c(\beta-x_0)}}{\beta-x_0} \right).
\ea
The fit involves only 4 parameters but leaving $\omega_{\bf p}$ free led to poorly constrained fits. Therefore, we fix the value of $\omega_{\bf p}$ to the prediction of \eqref{eq:dispersion_relation} with $u=u_m$ and ${\bf p} = (0,0,2\pi n/L)$ to see whether the data can be described in this way. Results are shown in \tabref{tab:u_m}. In view of the $\chi^2$-values, the data is consistent with this scenario. Moreover, the chiral effective theory makes a prediction for the residue of the retarded correlator at $\omega=\omega_{\bf p}$ reading (see App. B of \cite{ourref2})
\be
\rho^A(\omega, {\bf p}) = \underbrace{f^2_\pi (m^2_\pi + {\bf p}^2)}_{\text{Res}(\omega_{\bf p})} \delta(\omega^2-\omega^2_{\bf p}) + ...
\ee
which can be written in terms of $A_1({\bf p})$ and a new parameter $b({\bf p})$ which parametrizes the deviation with respect to the chiral prediction
\ba
\text{Res}(\omega_{\bf p}) &=& 2 A_1({\bf p}) \omega_{\bf p} \sinh(\omega_{\bf p}\beta/2)\\
&=& f^2_\pi (m^2_\pi + {\bf p}^2) (1 + b({\bf p})).
\ea
In addition, we rescale the parameter $A_2({\bf p}) \to \tilde{A}_2 = A_2({\bf p})/{\bf p}^2$ (neglecting the quark mass which is parametrically small) whose natural value is 1. It turns out that indeed the parameter $b{(\bf p})$ is small for $n=1$ and the values of $\tilde{A}_2$ are of order one adding confidence to our description.

\begin{table}[h]
\centering
\begin{tabular}{c|c|c|c|c|c|c|c}
$n$ & $A_1/T^3$ & $\omega_{{\bf p}_n}/T$ & $\tilde{A}_2$ 
 & $c/T$ & $\text{Res}(\omega_{{\bf p}_n})/T^4$ & $b$ & $\chi^2/{\rm d.o.f}$ \\ \hline 
1 & $2.95(4)\ten{-1}$ & $2.19(3)$ & $1.78(8)$ & $6.7(3)$ & $1.72(6)$ & $-0.08(3)$ & $0.06$ \\
2 & $1.40(5)\ten{-1}$ & $3.73(6)$ & $1.26(2)$ & $6.1(1)$ &$3.3(2)$ & $-0.39(4)$ & $0.15$ \\
3 & $4.9(3)\ten{-2}$ & $5.40(9)$ & $1.19(1)$ & $7.7(1)$ & $3.9(5)$ & $-0.65(4)$ & $0.35$ \\
4 & $1.7(2)\ten{-2}$ & $7.1(1)$ & $1.15(1)$ & $9.67(9)$ & $4.21(7)$ &$-0.78(3)$& $0.49$ \\
5 & $4(1)\ten{-3}$ & $8.8(1)$ & $1.12(1)$ & $11.7(1)$ &$3(1)$ &$ -0.89(3) $& $1.04$ \\
\hline \hline
\end{tabular}
\caption{Results of fits to the axial-charge density correlator at non-vanishing momentum ${\bf p}_n$.
All errors quoted are statistical, and
all renormalization factors are included. The quantity $\omega_{\bf p}/T$ is not a fit parameter;
rather it is set to the value predicted by \protect\eqref{eq:dispersion_relation} with $u(T)=u_m=0.74(1)$.}
\label{tab:u_m}
\end{table}

\section{Backus-Gilbert method on $G_A(x_0, T, {\bf p})$}
\begin{figure}[ht]
\begin{center}
\includegraphics[width=0.50\textwidth]{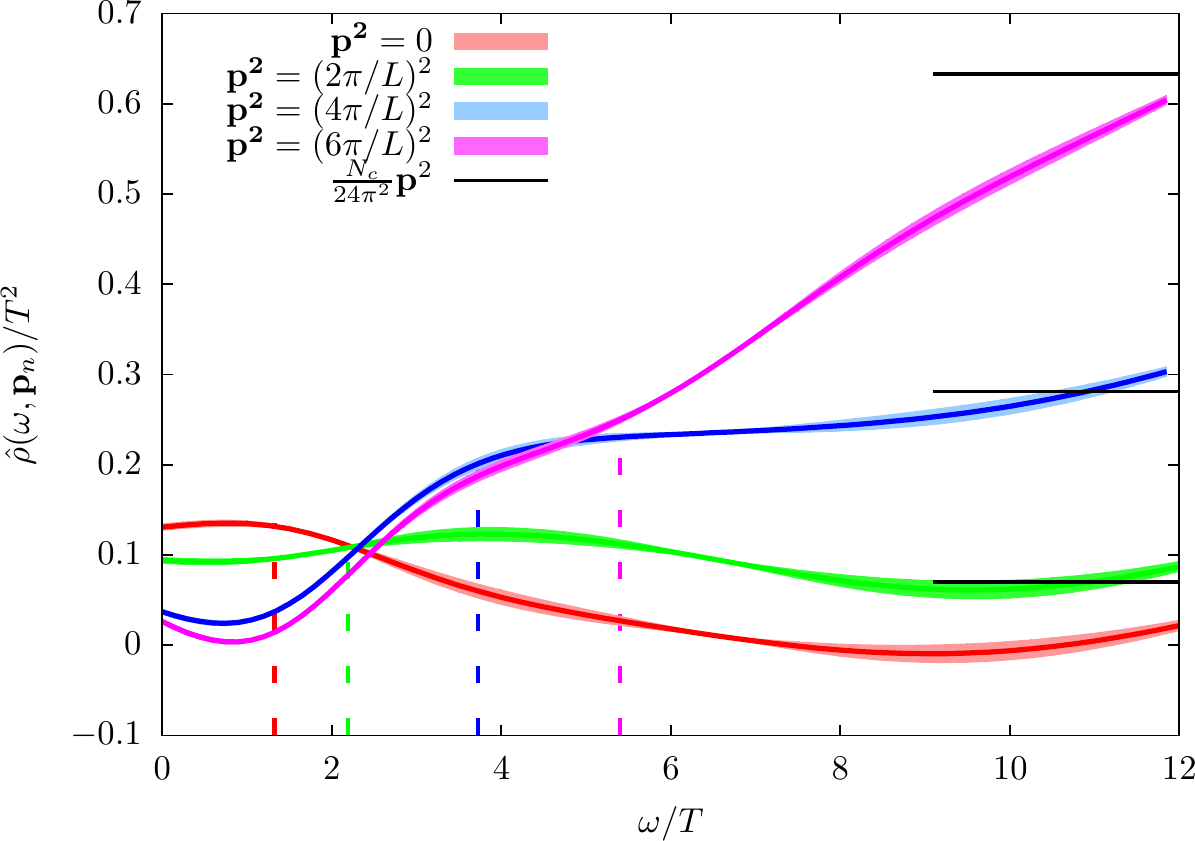}
\end{center} 
\caption{Estimators $\hat{\rho}(\omega, {\bf p}_n)/T^2$ for $n=0,1,2,3$
  together with its error shown as a band ($\lambda = 0.002$). The vertical colored dashed
  lines correspond to the locations of the expected positions of the
  poles $\omega_{\bf p}$ according to \protect\eqref{eq:dispersion_relation} with
  $u(T) = u_m$. The black horizontal lines correspond to the
  treelevel asymptotic values of $\rho^A(\omega, {\bf p})$. All renormalization
  constants have been taken into account. Dimensionful quantities
  have been made dimensionless by the appropriate power of $T=1/24a$.}
\label{fig:BG}
\end{figure}

\begin{figure}[h]
\begin{center}
\includegraphics[width=0.48\textwidth]{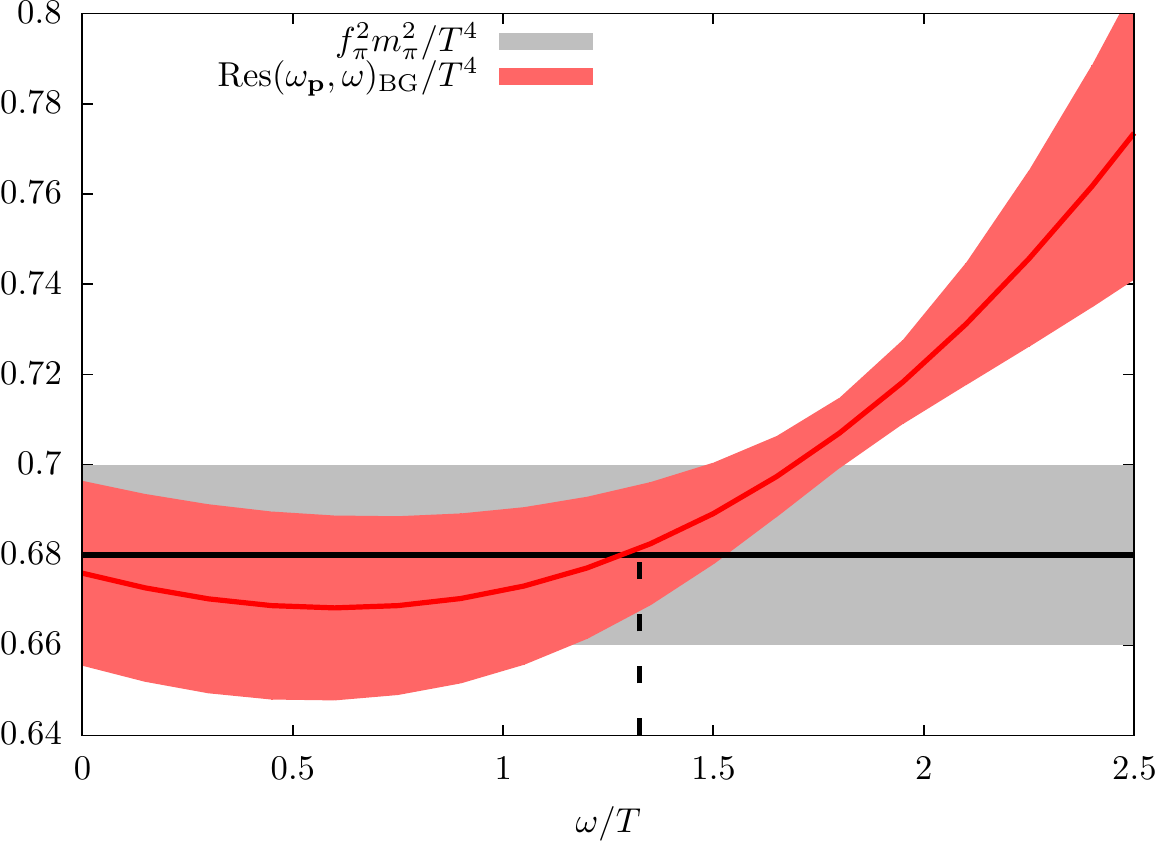}
\includegraphics[width=0.48\textwidth]{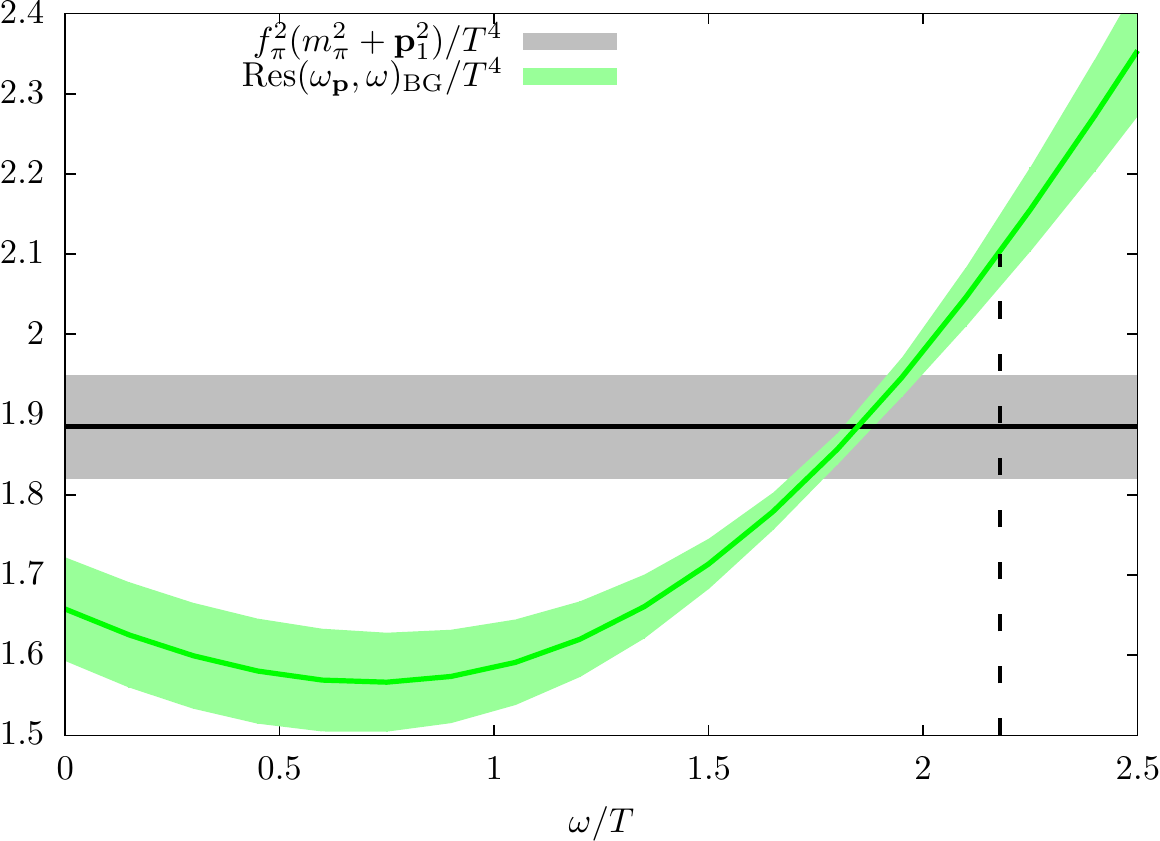}
\end{center}
\caption{The effective residue $\text{Res}(\omega_{\bf p}, \omega)_\text{BG}$ as
  defined in \protect\eqref{eq:res_BG}. Left: No momentum induced, ${\bf
    p}=0$. Right: One unit of momentum induced, ${\bf p}_1 =
  (0,0,2\pi/L)$. The grey band is the expectation in terms of
  screening quantities. All renormalization factors are
  included. The errors arise from the statistical uncertainty. The values 
of $\omega_{\bf p}$ are indicated by dashed vertical lines.}
\label{fig:res}
\end{figure}

The Backus-Gilbert method (see \cite{backusapp1,backusapp2} for more recent studies) is a technique for inverting integral equations like \eqref{eq:spectral_def} in order to directly obtain information on $\rho^A(\omega , {\bf p})$. The idea is to define an estimator 
\be
\hat{\rho}(\bar\omega, {\bf p}) = \int^\infty_0 d\omega \;\hat{\delta}(\bar\omega, \omega)\left( \frac{\rho_A(\omega,{\bf p})}{\tanh(\omega \beta/2)}\right)
\label{eq:hat_rho}
\ee
where $\hat{\delta}(\bar{\omega}, \omega)$ is called the resolution function which can be written in terms of a set of priori unknown coefficients $q_i(\bar{\omega})$ and the kernel functions $K_i(\omega) = \frac{\cosh(\omega(\beta/2-x^i_0))}{\cosh(\omega \beta/2)}$. It is a smooth function peaked around $\bar{\omega}$ which "filters" the true spectral function. The coefficients $q_i(\bar{\omega})$ are determined by minimizing its width subject to the condition that the area is normalized to 1. During the process the matrix $W$ need to be inverted, 
\be
W_{ij}(\bar{\omega}) = \lambda \left[\int^{\infty}_0 d\omega K_i(\omega)(\omega-\bar{\omega})^2 K_j(\omega)\right] + (1-\lambda)Cov(G_i,G_j), \qquad 0<\lambda < 1
\ee
where $Cov(G_i,G_j)$ denotes the $ij$-element of the covariance matrix of $G_A(x_0, T, {\bf p})$. The parameter $\lambda$ controls the trade off between stability and resolution power. For values of $\lambda$ close to 1, the matrix is almost singular and the error on $\hat{\rho}(\bar{\omega})$ rapidly increases. Reducing the value of $\lambda$ improves stability at the cost of deteriorating the frequency resolution. Finally the solution reads
\be
\hat{\rho}(\bar{\omega}, {\bf p}) = \sum_{i} q_i(\bar{\omega}) G_A(x^i_0, T, {\bf p}).
\ee
Notice that $\hat{\rho}(\bar{\omega}, {\bf p})$ is a completely model independent estimator in the sense that no ansatz was needed for its calculation. \figref{fig:BG} shows the estimators for different values of ${\bf p}$ and one already sees the good agreement in the high-frequency region with tree-level predictions. 
We can further check the agreement of $\text{Res}(\omega_{\bf p})$ with the chiral prediction by making use of $\hat{\rho}(\bar{\omega}, {\bf p})$. Assuming a $\delta$-type excitation in $\rho^A$ we can write
\be
\text{Res}(\omega_{\bf p}, \omega)_\text{BG}= \frac{2\omega_{\bf p}
 \tanh(\omega_{\bf p} \beta/2)\hat{\rho}(\omega, {\bf p})}{\hat{\delta}(\omega,\omega_{\bf p})}
\label{eq:res_BG}
\ee
where we use $\omega_{\bf p}$ as input according to \eqref{eq:dispersion_relation} with $u(T) = u_m$. This defines a function of $\omega$. The natural choice where $\text{Res}(\omega_{\bf p}, \omega)_\text{BG}$ is expected to be the best estimator for the true residue is at $\omega \approx \omega_{\bf p}$. By looking at \figref{fig:res} we see that approximately around this value the curve intercepts the grey band which represents the chiral prediction and therefore confirms the validity of \eqref{eq:dispersion_relation} at least up to ${\bf p} = 400\text{MeV}$ ($n=1$). 

\vspace{-.5cm}
\section{Conclusions}
We have successfully tested the dispersion relation of \eqref{eq:dispersion_relation} for the pion quasiparticle by performing direct fits to $G_A(x_0, T, {\bf p})$ (alternatively applying the Backus-Gilbert method) using the value of the parameter $u$ determined at zero momentum. \tabref{tab:screening} indicates that the pion mass "splits" at finite temperature into a lower pole mass and a higher screening mass. Our findings are in qualitative agreement with two-loop calculation in ChPT (see eg. \cite{schenk,toublan}). Both suggest a violation of boost invariance due to the presence of the medium. Although $N_\tau = 24$ is at the current state of the art concerning thermal ensembles, finite volume effects as well as lattice artifacts should be investigated. Our plans for the future are to simulate at the physical value of the light quark mass and/or to include the strange quark.

\acknowledgments
{We acknowledge the use of computing
  time for the generation of the gauge configurations on the \emph{JUGENE} and \emph{JUQUEEN}
  computers of the Gauss Centre for Supercomputing located at
  Forschungszentrum J\"ulich, Germany, allocated through the John von Neumann Institute for Computing (NIC) within project HMZ21. This work was supported by the
  \emph{Center for Computational Sciences in Mainz} as part of the
  Rhineland-Palatinate Research Initiative and by the DFG grant ME
  3622/2-1 \emph{Static and dynamic properties of QCD at finite
    temperature}.
}

\end{document}